\documentclass[aps,pra,twocolumn,amsfonts,showpacs,longbibliography,superscriptaddress]{revtex4-2}
\usepackage{verbatim,epsfig,amsmath,amssymb,bm,epsf,graphicx,psfrag,bbold,amsthm,amsfonts}
\usepackage[toc,page,titletoc]{appendix}
\usepackage[bottom]{footmisc}
\usepackage{hyperref}
\usepackage[capitalize]{cleveref} 
\hypersetup{
    colorlinks=true,
    linkcolor=black,
    citecolor=blue,
    filecolor=black,
    urlcolor=blue,
}

\usepackage{enumitem}
\usepackage{framed}
\usepackage{mathrsfs}
\usepackage{esint}
\setlist{nosep}
\usepackage{color}
\usepackage[utf8]{inputenc}
\usepackage{float}
\usepackage{natbib}
\usepackage{tikz-cd}
\usepackage{leftidx}
\usepackage[normalem]{ulem}
\usepackage[caption=false]{subfig}
\usepackage{notes2bib}

\newcommand{\moy}[1]{\langle #1 \rangle}

\newcommand\cT {{\mathcal{T}}}
\newcommand\cR {{\mathcal{R}}}
\newcommand\cZ {{\mathcal{Z}}}

\newcommand{\cP}{\mathcal{P}}

\newcommand\bZ {{\mathbb Z}}

\newcommand{\ztzt}{\mathbb{Z}_2 \times \mathbb{Z}_2}
\newcommand{\ztwo}{\mathbb{Z}_2}

\theoremstyle{plain}

\usepackage{cancel}
\theoremstyle{definition}

\theoremstyle{remark}

\begin{document}
\title{\large Classical Origins of Landau-Incompatible Transitions}
	\author{Abhishodh Prakash}
    \email{abhishodh.prakash@physics.ox.ac.uk}
    \altaffiliation{(he/him/his)}
	\affiliation{Rudolf Peierls Centre for Theoretical Physics, University of Oxford, UK}
        \affiliation{Harish-Chandra Research Institute, Prayagraj (Allahabad), India}
 	\author{Nick G. Jones}
    \email{nick.jones@maths.ox.ac.uk}
        \altaffiliation{The published version of this article is Phys. Rev. Lett. 134, 097103 (2025).\\ \href{ https://doi.org/10.1103/PhysRevLett.134.097103}{ https://doi.org/10.1103/PhysRevLett.134.097103}}
	\affiliation{St John’s College and Mathematical Institute, University of Oxford, UK}
	
	\begin{abstract}
Continuous phase transitions where symmetry is spontaneously broken are ubiquitous in physics and often found between `Landau-compatible' phases where residual symmetries of one phase are a subset of the other. However, continuous `deconfined quantum critical' transitions between Landau-incompatible symmetry-breaking phases are known to exist in certain quantum systems, often with anomalous microscopic symmetries.  In this Letter, we investigate the need for such special conditions. We show that Landau-incompatible transitions can be found in a family of well-known classical statistical mechanical models with anomaly-free symmetries, introduced by Jos\'{e}, Kadanoff, Kirkpatrick and Nelson (Phys. Rev. B 16, 1217). The models are anisotropic deformations of the classical 2d XY model labelled by a positive integer $Q$. For a range of temperatures, even $Q$ models exhibit two Landau-incompatible partial symmetry-breaking phases and a direct transition between them for $Q \ge 4$. Characteristic features of deconfined quantum criticality, such as enhanced symmetries and melting of charged defects are easily seen in a classical setting. For odd $Q$, and corresponding temperature ranges, two regions of a single partial symmetry-breaking phase appear, split by a stable `unnecessary critical' line. We discuss experimental systems that realize these transitions and present anomaly-free quantum models that also exhibit similar phase diagrams. 	
\end{abstract}

	\maketitle
 \noindent   Spontaneous symmetry breaking (SSB) underpins several important physical phenomena, from the development of long-range orders in matter to endowing mass to fundamental particles~\cite{Gross1996SSB}. The simplest setting for SSB is when a phase of matter, classical or quantum, with a vacuum invariant under a symmetry group $G$ undergoes a phase transition to produce multiple vacua, each of which preserves only a subset of the original symmetries $H \subset G$.  If such a phase transition is continuous, it can be described within the Landau-Ginzburg-Wilson-Fisher (LGWF) framework using a local order parameter field.  About twenty years ago, the nature of exotic direct transitions between incompatible SSB quantum phases, where the vacuum symmetries of neither phase could be identified as a subset of the other, was investigated in two-dimensional quantum systems~\cite{OGDQC}. Although such transitions had appeared in earlier studies~\cite{LSMOG}, Ref.~\cite{OGDQC} recognized that they could not be framed within the LGWF paradigm in terms of order parameter fields. Instead, they were naturally formulated using gauge fields, which are hidden from sight in the ordered phases but appear at the transition. This prompted the moniker `deconfined quantum criticality' (DQC)~\cite{OGDQC,Senthil2023DQCReview}. 

\begin{figure}[!htp]
    \centering
    \includegraphics[]{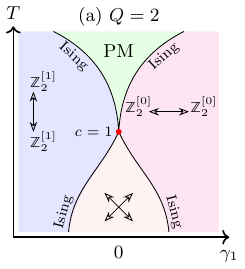}\hspace{-1.2em}
    \includegraphics[]{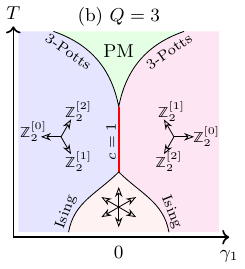}
    \includegraphics[]{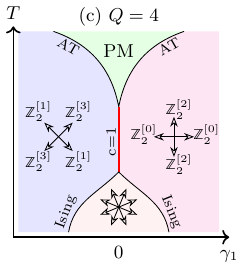} \hspace{-1.2em}
    \includegraphics[]{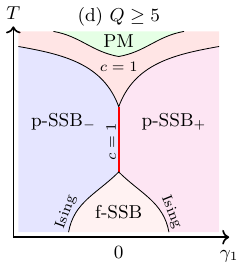}
    \caption{Phase diagrams for the JKKN Hamiltonian shown in \cref{eq:Hclassical} for $Q \ge 2$ and fixed $\gamma_2 <0$. The arrows represent the different vacua characterized by the expectation value $\moy{\theta_j}$. The residual symmetry for each vacuum is shown following the notation in \cref{eq:ztwoalpha}. For even $Q\ge 4$, the red line along $\gamma_1=0$, represents a direct transition between Landau-incompatible partial symmetry-breaking (p-SSB) phases. For odd $Q \ge 3$, it represents an unnecessary critical transition. The labels Ising, Ashkin-Teller (AT), 3-state Potts, and compact boson ($c=1$) indicate the conformal field theory describing the transition.  At high temperatures we have a paramagnetic (PM) phase. }
    \label{fig:Phasediagrams}
\end{figure}
    
What physical settings can give rise to such Landau-incompatible transitions? Low-dimensional examples~\cite{DQC1d_Robertsetal_PhysRevB.99.165143,DQC1d_Mudryetal_PhysRevB.99.205153,DQC_1D_SPT_Zhang_PhysRevLett.130.026801,Chatterjee23}   and descriptions using non-linear sigma models~\cite{SenthilFisher_DQC_SigmaPhysRevB.74.064405} have clarified that deconfined gauge fields are not essential. However, quantum effects are believed to play a necessary role in DQC~\cite{Singh2010DQCQuantum}, especially when viewed from the role played by Berry phases, and most examples of DQC occur under Lieb-Schultz-Mattis conditions~\cite{LSMOG}, when microscopic symmetries are \emph{anomalous}~\cite{Wangetal_DQC_2017_PhysRevX.7.031051,MetlitskiThorngrenDQC_PhysRevB.98.085140}.

In this Letter, we show that these conditions are \emph{not} necessary and Landau-incompatible transitions can, in fact, be found even in ordinary classical statistical mechanical systems with anomaly-free symmetries. We demonstrate this using a well-known family of models introduced by Jos\'{e}, Kadanoff, Kirkpatrick and Nelson (JKKN)~\cite{JoseKirkpatrickPhysRevB.16.1217} obtained by perturbing the 2d classical XY model by on-site anisotropies labelled by a positive integer $Q$. For even $Q\ge 4$, the phase diagram includes a  direct phase transition between two Landau-incompatible partial symmetry-breaking phases. This transition displays all notable characteristics of DQC, including the appearance of an enhanced symmetry that rotates between the order parameters of the Landau-incompatible phases, and the melting of charged defects. Interestingly, neutral defects and isolated charges also independently exist and can condense to produce Landau-compatible phase transitions. For odd $Q$, we find an exotic second-order `unnecessary critical' line separating two regions of single phase that does not represent a genuine phase transition~\cite{BiSenthilPhysRevX.9.021034,AP_UC_PhysRevLett.130.256401} but is nevertheless stable. To our knowledge, this is the first case of unnecessary criticality identified in a classical model.  We discuss experimental systems that exhibit these phenomena and also present a family of quantum models with anomaly-free symmetries that exhibit similar phase diagrams. Our results show that exotic transitions beyond the LGWF paradigm are more abundant than previously believed. 
  
\medskip
	\noindent \emph{\textbf{Models and phase diagram:}} 
 Let us consider the JKKN model~\cite{JoseKirkpatrickPhysRevB.16.1217}, a classical statistical mechanical system of planar rotors $\theta_j \sim \theta_j +2\pi$ located on the vertices of any two-dimensional lattice. The Hamiltonian is
 \begin{align}
	H = -\sum_{\moy{j,k}} \cos \left(\theta_j - \theta_k \right) - \sum_{\ell = 1,2,\ldots} \gamma_\ell \sum_j  \cos \left(\ell Q \theta_j \right) \label{eq:Hclassical}.
\end{align}
Here, $j$ labels the vertices and $\moy{j,k}$ the edges of the lattice. The relevant symmetry group of the system is generated by a rotation $\cR$ and a reflection $\cT$ acting as 
\begin{align}
\cR :\theta_j \mapsto \theta_j + {2 \pi}/{Q} , \qquad~\cT: \theta_j \mapsto -\theta_j. \label{eq:symmetries}
\end{align}
The two generators do not commute but satisfy  
\begin{align}
   \cT \cR = \cR^{-1} \cT,~\cR^Q = \cT^2 = 1,
\end{align}
and form the non-abelian dihedral group $D_{2Q} \cong \bZ_Q \rtimes \ztwo$.  We will be interested in the equilibrium phase diagram of \cref{eq:Hclassical},  varying  the temperature $T = \beta^{-1}$ and $\gamma_1$ close to $\gamma_1 = 0$, while keeping all other couplings fixed, the most important being $\gamma_2$. This is shown schematically in \cref{fig:Phasediagrams} for $\gamma_2 <0$ and all $|\gamma_{\ell}|$ kept small \cite{supp}. Let us summarize its main aspects~\cite{JoseKirkpatrickPhysRevB.16.1217,LECHEMINANT_SelfDualSineGordon_2002502}: 

\smallskip 

\begin{enumerate}[wide, labelindent=0pt]
    \item  At low temperatures and $\gamma_2 <0$, we obtain an ordered phase with full symmetry breaking and $2Q$ vacua (abbreviated f-SSB). For $\gamma_2>0$, this becomes a first-order line separating partial SSB regions described below  \cite{supp}.
    \item For a range of intermediate temperatures, we obtain two regions with partial SSB (abbreviated p-SSB$_\pm$ for $\gamma_1 >0$ and $\gamma_1<0$ respectively), each containing $Q$ vacua. These are separated from the f-SSB phase by an Ising transition for any $Q$~\cite{DELFINO1998675}.  For even $Q$, p-SSB$_\pm$ represent two distinct Landau-incompatible SSB phases, whereas for odd $Q$, they correspond to the same phase. 
    \item For $Q\ge 3$, the two partial SSB regions are separated by a critical line at $\gamma_1 = 0$ which is of prime interest. For even $Q$, this is a direct, stable Landau-incompatible transition. For odd $Q$, this line represents `unnecessary criticality'~\cite{BiSenthilPhysRevX.9.021034,AP_UC_PhysRevLett.130.256401} and is expected to terminate  under appropriate strong perturbation.
    \item At high temperatures, we get a disordered paramagnetic phase (PM) that restores all symmetries. This is separated from the partial SSB phases by a direct transition belonging to the Ising, 3-state Potts and Ashkin-Teller universality classes (or their symmetry-enriched variants~\cite{VerresenSEC,AP_SEC_PhysRevB.108.245135,supp}) for $Q=2,3,4$ respectively~\cite{LECHEMINANT_SelfDualSineGordon_2002502}, and by an intermediate gapless phase  for $Q \ge 5$~\cite{JoseKirkpatrickPhysRevB.16.1217}.
\end{enumerate}

The phase diagrams in \cref{fig:Phasediagrams} are determined by replacing \cref{eq:Hclassical} by an effective gaussian continuum theory~\cite{JoseKirkpatrickPhysRevB.16.1217,LECHEMINANT_SelfDualSineGordon_2002502} via a duality transformation  \`{a} la Villain~\cite{Villain1975theory}
\begin{align}
    S \approx  \int d^2x \left[ \frac{ \left( \nabla \phi\right)^2}{8 \pi^2 \beta}   - h  \cos (\phi)  -  \sum_{\ell}  \gamma_\ell \cos \left(\ell Q \theta\right)  \right], \label{eq:S}
\end{align}
and keeping track of the relevance (in the renormalization group sense) of the scaling operators  $\cos (\phi)$ and $\cos( \ell Q \theta)$  \cite{supp,JoseKirkpatrickPhysRevB.16.1217}. Recall that a scaling operator $\mathscr{O}$ is relevant when its scaling dimension $[\mathscr{O}]$ is smaller than the spatial dimension (two in our case) for such classical statistical mechanical systems~\cite{Ginsparg1988applied}.  The term $h \cos (\phi)$ is included as a regulator in the Villain procedure. Close to the fixed point described by the conformal field theory (CFT), the scaling dimensions are determined by a single stiffness parameter $K_{\rm eff}$, as~\cite{DiFrancesco:1997nk,JoseKirkpatrickPhysRevB.16.1217,supp} 
\begin{equation}
    [\cos (\phi)] = \pi K_{\rm eff},~ [\cos( \ell Q \theta)] = \frac{\ell^2 Q^2}{4 \pi K_{\rm eff}}. \label{eq:Keff}
\end{equation}
While the exact relationship between $K_{\rm eff}$ and the microscopic parameters $T,h,\gamma_\ell$ cannot be determined exactly, 
we see that for $Q\ge 4$, there exists a regime $Q^2/(8\pi)< K_{\rm eff} <Q^2/(2\pi)$ where $\cos (Q \theta)$ is the only relevant symmetry-allowed operator. Tuning this away by setting $\gamma_1 = 0$ produces a critical state corresponding to the Landau-incompatible transition or unnecessary critical line. Important parts of the phase diagrams in \cref{fig:Phasediagrams} have already been explored in previous work (see Refs.\cite{JoseKirkpatrickPhysRevB.16.1217,LECHEMINANT_SelfDualSineGordon_2002502}). Our main focus will be on symmetry properties, Landau-(in)compatible nature of transitions, unnecessary criticality, and the distinction between even and odd $Q$. These aspects have not been investigated previously, to the best of our knowledge. 

\medskip

\noindent \emph{\textbf{Residual symmetries, Landau (in)compatibility:}}
Let us understand the nature of the partial symmetry breaking regions (p-SSB$_\pm$) which are realized for $\gamma_1>0$ and $\gamma_1<0$, respectively, at intermediate temperatures, where the only relevant operator in \cref{eq:S} is $\cos (Q \theta)$. Both have $Q$ vacua $\vartheta^\pm_1,\ldots,\vartheta^\pm_Q$ characterized by the vacuum expectation value  $ \moy{\theta_j} = \vartheta^\pm_n$ with 
\begin{align}
 \vartheta^+_n = {2 \pi n}/{Q},~   \qquad \vartheta^-_n = {(2n+1) \pi }/{Q}.  \label{eq:Qvacua}
\end{align}
The symmetries shown in \cref{eq:symmetries} act on the vacua as 
\begin{align}
    \cR: \vartheta^\pm_n \mapsto \vartheta^\pm_{n+1},~ \cT: \begin{pmatrix}
        \vartheta_n^+\\
        \vartheta_n^-
    \end{pmatrix} \mapsto \begin{pmatrix}
        \vartheta_{-n}^+\\
        \vartheta_{-n-1}^-
    \end{pmatrix} \label{eq:pSSB symmetry}.
\end{align}
 
Since both regions have the same number of vacua, only two possibilities exist: (i) the regions are distinct phases that are Landau-incompatible or (ii) they correspond to the same phase. To clarify which, we need to determine the residual symmetries of each vacuum,  $\mathscr{I}\left(\vartheta^\pm_n\right)$, in both regions. Using \cref{eq:pSSB symmetry}, we see that the vacua transform into each other under the discrete rotations $\cR$ but preserve a specific $\ztwo$ subgroup generated by reflection $\cT$ followed by a certain number of rotations $\cR$. To distinguish between various $\ztwo$ groups, we define the following notation:
\begin{align}
    \ztwo^{[\alpha]} \equiv \{ 1,\cR^\alpha \cT \} \label{eq:ztwoalpha}
\end{align}
with $\alpha = \alpha + Q$ identified. Using these, we get 
\begin{align}
    \mathscr{I}(\vartheta^+_n) =   \ztwo^{[2n]},~\mathscr{I}(\vartheta^-_n) =  \ztwo^{[2n+1]}, \text{ for}~n=0,\ldots,Q-1. \label{eq:residual_general}
\end{align}
\cref{fig:Phasediagrams} shows the residual symmetries for $Q=2,3,4$. For even $Q$, these are distinct for p-SSB$_\pm$ and the vacua are invariant under $\cT$ followed by even (odd) $\cR$  rotations for $\gamma_1 >  0 ~(\gamma_1 <0)$. The p-SSB$_\pm$ phases are detected, respectively, by the following order parameters,
\begin{align}
 \mathcal{E}_+ = \cos \left({Q\theta}/{2}\right),~  \mathcal{E}_- =\sin \left({Q\theta}/{2}\right).\label{eq:SSBpm order parameters}
 \end{align}
 There is no way to identify the residual symmetries of the vacua of one phase with subsets of those of the other, and therefore the phases are distinct and Landau-incompatible. We see the advertised direct transition between them along $\gamma_1=0$ for a range of temperatures with continuous critical exponents described by the gaussian CFT that \cref{eq:S} flows to. 
 
 For odd $Q$ on the other hand, the residual symmetries for both p-SSB$_\pm$ are identical and detected by the same order parameter, 
 \begin{align}
 \mathcal{O} = \cos \big({(Q-1)\theta}/{2}\big).\label{eq:SSBodd order parameters}
 \end{align}
The vacua on both sides can be identified as follows
\begin{align}
    \mathscr{I}\left(\vartheta^+_{n + \frac{Q+1}{2}}\right) = \mathscr{I}\left(\vartheta^-_n\right) \implies \vartheta^+_{n + \frac{Q+1}{2}} \cong \vartheta^-_n.
\end{align}
We conclude that both belong to the same phase and that there should exist a path where the vacua $\vartheta^+_{n + \frac{Q+1}{2}}$ and $\vartheta^-_n$ can be smoothly connected without encountering a phase transition. Examples of explicit paths that connect the two regions are sketched in the end matter. Thus, for odd $Q$, the critical line along $\gamma_1=0$ represents `unnecessary criticality'~\cite{BiSenthilPhysRevX.9.021034,AP_UC_PhysRevLett.130.256401} which does not correspond to a genuine transition separating distinct phases but is nevertheless stable and reached by tuning a single relevant parameter. To the best of our knowledge all known instances of unnecessary criticality~\cite{AnsudoRoschPhysRevB.75.144420,MoudgalayaPollmanPhysRevB.91.155128,AP_UC_PhysRevLett.130.256401,Verresen2021quotient,BiSenthilPhysRevX.9.021034,YuchiTerminablePhysRevLett.132.136501} have been observed in quantum mechanical systems~\cite{WangWenWittenPhysRevX.8.031048,APUnwindingPhysRevB.98.125108,APUnwindingfermionPhysRevB.103.085130},  and ours is the first example in a classical SSB setting.

For completeness, let us consider the remaining phases and transitions in \cref{fig:Phasediagrams}. The full symmetry breaking phase (f-SSB) appears when $\cos (2Q\theta)$ becomes relevant along the $\gamma_1 = 0$ line at low temperatures. This is detected by the order parameter 
\begin{equation}
    \mathcal{F} = \sin (Q\theta) \label{eq:SSBallorderparameter}
\end{equation}
and has $2Q$ vacua that break all symmetries.  The transition between p-SSB$_\pm$ and f-SSB is Landau-compatible and, when continuous, belongs to the Ising universality class~\cite{supp,DELFINO1998675}. Finally, at large $T$ all symmetries are restored when $\cos(\phi)$ becomes relevant along the $\gamma_1 = 0$ line to produce a disordered paramagnet (PM). For $Q \le 4$, this transition is direct and also Landau-compatible; the universality class depends on $Q$~\cite{supp,LECHEMINANT_SelfDualSineGordon_2002502}.

\medskip
\begin{figure*}
    \centering
    \begin{tabular}{cccc}
    \begin{tabular}{c}
       \includegraphics[]{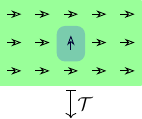} \\
       \includegraphics[]{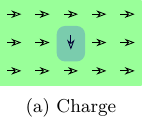}
    \end{tabular}
       &     \begin{tabular}{c}
       \includegraphics[]{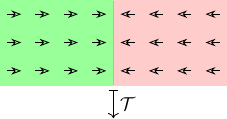} \\
       \includegraphics[]{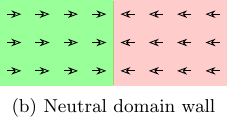}
    \end{tabular}  &
            \begin{tabular}{c}
       \includegraphics[]{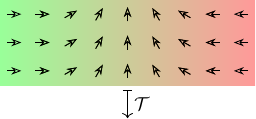} \\
       \includegraphics[]{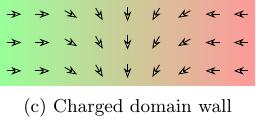}
    \end{tabular} &     \begin{tabular}{c}
     \vspace{1.1em}  \includegraphics[]{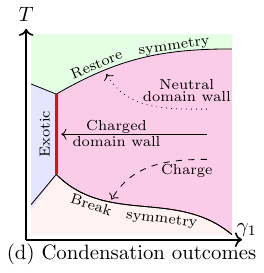}\\
    \end{tabular}
    \end{tabular}
   \caption{Schematic representation of various excitations in the pSSB$_+$ phase and their symmetry transformation: (a) local charges, (b) neutral and (c) charged domain walls. The latter can be regarded as a bound state of the former two. The transition triggered by condensing each excitation is shown in (d). The neutral domain wall does not carry symmetry charges and its proliferation restores all symmetries, whereas condensing the charge breaks all symmetries. Melting the charged domain walls produces the exotic transitions discussed in this work: Landau-incompatible transitions for even $Q$ which breaks some symmetries while restoring others, and unnecessary criticality for odd $Q$.}
    \label{fig:DomainWall}
\end{figure*}

\noindent \emph{\textbf{Enhanced symmetries, charged defects:}} We now study two prominent aspects of DQC transitions in our classical models. The first is the appearance of enhanced symmetries~\cite{Senthil2023DQCReview} involving rotations between order-parameters of the two Landau-incompatible phases that the DQC line separates. This is readily seen for our model. For even $Q$, we can use the order parameters of p-SSB$_\pm$ phases shown in \cref{eq:SSBpm order parameters} to define the two-component unit vector,
\begin{equation}
    \hat{n} = \begin{pmatrix}
        \mathcal{E}_+,\mathcal{E}_-
    \end{pmatrix}= \begin{pmatrix} 
        \cos\left(Q \theta/2 \right),
        \sin\left(Q \theta/2 \right)
    \end{pmatrix}. \label{eq:nhat}
\end{equation}
Along the direct Landau-incompatible transition, the $D_{2Q}$ symmetry of the JKKN model is enhanced to the $O(2)$ symmetry of the XY model whose order parameter is $\hat{n}$. This is generated by a full $\theta$ rotation which transforms the two components of $\hat{n}$, $\mathcal{E}_\pm$ into each other.

The second aspect is the physical picture for the onset of DQC being the proliferation of charged defects that prevents the restoration of symmetries~\cite{Senthil2023DQCReview}. This is also clearly seen in our model. In the vicinity of the DQC transition along $\gamma_1 = 0$, the relevant excitations are smooth interpolations between vacua with the same residual symmetries. The domain walls resulting from this interpolation are charged under the residual $\ztwo$ symmetry. For illustration, let us focus on $\gamma_1>0$ where, for even $Q$, two of the vacua of the resulting p-SSB$_+$ phase are $\moy{\theta_j} = 0 \text{ and }\pi$ with the same residual symmetry, $\ztwo^{[0]}=\{1,\cT\}$. If we create a smooth interpolation between these vacua as shown in \cref{fig:DomainWall}(c), we see that the resulting domain wall transforms under $\cT$ and thus carries charge. Furthermore, by evaluating the order parameter $\mathcal{E}_+$ on this configuration, we see that it vanishes on the domain wall, whereas the order parameter $\mathcal{E}_-$, which vanishes everywhere else, becomes non-zero at the domain wall. Thus, upon melting the p-SSB$_+$ domain walls, we get  p-SSB$_-$ order!

\medskip
\noindent \emph{\textbf{Experimental realizations:}} An outstanding challenge for DQC is the relative paucity of experimental platforms for its validation~\cite{Senthil2023DQCReview}. The results presented in this Letter open up avenues in classical systems, where Landau-incompatible transitions can be studied more easily. In fact, they have already been observed in several existing experimental systems~\cite{Kosterlitz_2016,Hrec_theoryPhysRevB.47.2333,Hrec_theoryPhysRevB.50.12692,Hreconstruction_PhysRevB.33.7906,Hreconstruction_GRIFFITHS1981671,RauRobert_XY_Fe,ROELOFS1982425_HQ_Expt,FeAu_PhysRevLett.62.206,FeAuRAU1990406,MoW_Hqexpt_PhysRevLett.38.1138}. For example, the $Q=4$ model of \cref{eq:Hclassical} describes the adsorption of hydrogen on the (100) surface of tungsten~\cite{Hrec_theoryPhysRevB.47.2333,Hrec_theoryPhysRevB.50.12692}. This system exhibits a structural transition~\cite{Hreconstruction_PhysRevB.33.7906,Hreconstruction_GRIFFITHS1981671} which is nothing but the Landau-incompatible transition shown in \cref{fig:Phasediagrams}(c). The same model also describes ultrathin deposits of iron on gold substrate~\cite{RauRobert_XY_Fe,ROELOFS1982425_HQ_Expt,FeAu_PhysRevLett.62.206,FeAuRAU1990406}.

\medskip
\noindent \emph{\textbf{Quantum models:}}
All important parts of the phase diagrams of \cref{eq:Hclassical} are qualitatively reproduced by the ground states of the quantum Hamiltonian
\begin{align}
	H &= -H_{\textrm{XXZ}} - h H_0 - \sum_{\ell = 1,2,\ldots} \gamma_{\ell} H_{ \ell Q}  \label{eq:Hquantum} \\
 \text{ where, } & H_{\text{XXZ}} =  \sum_j \left(  S^x_j S^x_{j+1} + S^y_j S^y_{j+1}+ \Delta  S^z_j S^z_{j+1}  \right) ,\nonumber \\
 H_{\ell Q} &= \sum_j \left( \prod_{l=0}^{\ell Q-1} S^+_{j+l} + \prod_{l=0}^{\ell Q-1} S^-_{j+l} \right), \text{ and } \nonumber \\
 H_0 &=\sum_j (-1)^j S^z_j \text{ or } \sum_j (-1)^j \left(  S^x_j S^x_{j+1} + S^y_j S^y_{j+1} \right). \nonumber 
\end{align}
$\vec{S} = \frac{1}{2}\vec{\sigma}$ are standard spin half angular momentum operators, $H_{\textrm{XXZ}}$ is the XXZ spin chain and $H_Q$ is a term that favours SSB. Both choices of $H_0$ favour disordered paramagnets that preserve all symmetries, but the second choice also produces a symmetry-protected topological (SPT) phase for one of the signs of $h$~\cite{AP_SEC_PhysRevB.108.245135}. The symmetries in \cref{eq:symmetries} are generated by $
	\cR = \prod_j \exp({\frac{2 \pi i}{Q} S^z_j })
$, $\cT$ is a time-reversal symmetry generated by complex conjugation in the $Z$ basis and the local order parameters corresponding to \cref{eq:SSBpm order parameters,eq:SSBodd order parameters,eq:SSBallorderparameter} are
\begin{align}
    & \mathcal{E}_\pm = \cP_\pm(Q/2),~ \mathcal{O} = \cP_+((Q-1)/2),~\mathcal{F} = \cP_-(Q)  \\
    & \text{where, } \cP_\pm(M) \equiv e^{ \frac{i \pi(1\mp1)}{4}}\left( \prod_{l=0}^{M-1} S^+_{j+l} \pm \prod_{l=0}^{M-1} S^-_{j+l} \right) \nonumber. 
\end{align}
This model can be bosonized to get the same field theory as in \cref{eq:S} for $|\Delta| <1$, up to renormalization of coupling constants. 

\medskip
\noindent \emph{\textbf{The role of anomalies:}}  The p-SSB$_{\pm}$ phases have two more distinct type of excitations. The first are charges shown in \cref{fig:DomainWall}(a) corresponding to local deviations from the $\gamma_1 \cos (Q\theta)$ minima, whose condensation further breaks symmetries and produces the transition to the f-SSB phase. The second are `hard' domain walls shown in \cref{fig:DomainWall}(b) that do not transform under the residual symmetries and are favoured at large values of $|\gamma_1|$. Proliferating these by increasing temperature restores all symmetries and drives the transition to the disordered PM. For small $|\gamma_1|$, charges are bound to neutral domain walls to produce the soft domain walls shown in \cref{fig:DomainWall}(c). As $\gamma_1 \rightarrow 0$, these bound states, rather than their constituents, melt to drive the Landau-incompatible transitions studied in this letter.

In several models exhibiting DQC~\cite{OGDQC,DQC_1D_SPT_Zhang_PhysRevLett.130.026801,DQC1d_Robertsetal_PhysRevB.99.165143}, the binding of charges to defects occurs kinematically, due to the \emph{anomalous} nature of underlying microscopic symmetries~\cite{Senthil2023DQCReview,DQC1d_Robertsetal_PhysRevB.99.165143,DQC1d_Mudryetal_PhysRevB.99.205153,DQC_1D_SPT_Zhang_PhysRevLett.130.026801}. Anomalies are exotic symmetry representations found in systems constrained by Lieb-Schultz-Mattis conditions~\cite{LSMOG,Hastings_LSM_PhysRevB.69.104431,Oshikawa_LSM_PhysRevLett.84.1535,AP_LSM2020elementary} or on the boundaries of SPT phases~\cite{ElseNayakCohomologyPhysRevB.90.235137,DQC_1D_SPT_Zhang_PhysRevLett.130.026801}. They forbid strictly on-site representations~\cite{ElseNayakCohomologyPhysRevB.90.235137}, present an obstruction to gauging~\cite{WenAnomalies_PhysRevD.88.045013,fradkin2021quantum} and disallow a trivial symmetry-preserving phase~\cite{MetlitskiThorngrenDQC_PhysRevB.98.085140,AP_LSM2020elementary}. The latter feature is reflected in the absence of neutral defects that can proliferate to form a trivial phase and permitting only charged ones which can condense to produce DQC~\cite{Senthil2023DQCReview}. 

All microscopic symmetries in \cref{eq:symmetries} are anomaly-free. This is verified by the presence of the symmetry-allowed operator $\cos (\phi)$ in \cref{eq:S} which produces a trivial phase---a sufficient condition for the absence of anomalies. What we have in our model is arguably a more pedestrian route for the binding of charges to defects---such a bound state may find itself energetically more favourable~\footnote{An analogous charge-binding transition and associated exotic critical phenomena in a 2d classical model were investigated in Refs~\cite{FendleyLamacraft_XYPhysRevLett.107.240601,FendleyChalkerXY_2017}}.  However, when $\cos (\phi)$ is irrelevant, a new continuous $\phi$ rotation symmetry emerges that is preserved by all remaining relevant operators~\footnote{The critical Ising model, which is the archetypal Landau-compatible transition, also has a mixed anomaly between the microscopic Ising symmetry and the emergent Kramers-Wannier symmetry~\cite{Zhang2023anomalies}. The key difference is that this mixed anomaly is explicitly broken by \emph{relevant} operators, whereas the same is broken by \emph{irrelevant} operators in our model. }  which has a mixed anomaly with the microscopic $\cR$ symmetry~\cite{MetlitskiThorngrenDQC_PhysRevB.98.085140,fradkin2021quantum}. This microscopic-emergent mixed anomaly may be said to stabilize the Landau-incompatible and unnecessary critical transitions. It is unclear if this is a necessary precondition. 

\medskip

\noindent \emph{\textbf{Outlook:}} We have investigated classical 2d statistical mechanical models hosting stable Landau-incompatible transitions and unnecessary criticality. These transitions are driven by the melting of charged defects and stabilized by a mixed anomaly between microscopic and emergent symmetries unbroken by relevant operators despite all microscopic symmetries being anomaly-free.  

Our work opens several lines of future investigation. An obvious one is whether we can find other classical models that exhibit similar phenomena~\footnote{Numerical evidence for a direct Landau-incompatible transition in an alternative classical model was presented in Ref.~\cite{Chatterjee23b}}, especially in higher dimensions.   The archetype DQC transition, between N\'{e}el to valence-bond-solid phases~\cite{OGDQC} has recently been shown to be first-order in nature~\cite{Sandviktakahashi2024so5multicriticalitytwodimensionalquantum}  and it would be interesting to find alternative, classical settings where a Landau-incompatible transition can be present between other phases.  For odd $Q$ models, it would be interesting to see how the unnecessary critical surface terminates. In \cite{AP_UC_PhysRevLett.130.256401,prakash2024chargepumpsboundarymodes}, it was argued that unnecessary criticality in quantum models is stabilized by the encircling states forming a non-trivial \emph{family} ~\cite{kitaev2019,ThorngrenDiabolicalPhysRevB.102.245113,HermeleGappedfamiliesPhysRevB.108.125147}. We expect this to be true for our classical model and it would be most interesting to explore this connection further. It would also be useful to further clarify if the stable mixed microscopic-emergent anomaly is a necessary condition for the exotic transitions studied here.

We have shown that exotic transitions can exist under relatively ordinary conditions within reach of existing experiments. It would be gratifying to validate this in more experimental setups. Finally, it would be illuminating to explore what other phenomena attributed to quantum fluctuations can have a classical origin.

 \medskip
 \noindent \emph{\textbf{Acknowledgments}}: The authors thank Ruben Verresen, Ryan Thorngren, Steve Simon, Paul Fendley, Sounak Biswas, Yuchi He and Saranesh Prembabu for useful discussions and the anonymous referees for insightful comments. A.P. was supported by the European Research Council under the European Union Horizon 2020 Research and Innovation Programme, Grant Agreement No. 804213-TMCS.

\bibliography{references}{}

  \renewcommand\appendixpagename{\centering \large \uline{End Matter}}
\begin{appendices}
\subsection*{\uline{Spontaneous symmetry breaking and Landau-compatibility}}\label{app:SSB}
\begin{figure}[!h]
    \centering
    \subfloat[$G = D_4 \cong  \ztwo \times \ztwo$]{\includegraphics[width=0.38\textwidth]{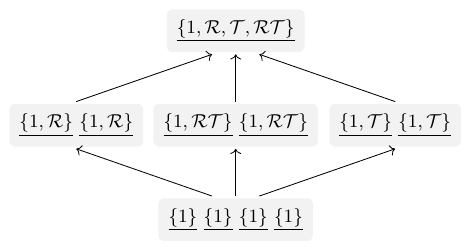}}  \\ 
    \subfloat[$G = D_8 \cong \bZ_4 \rtimes \ztwo$]{\includegraphics[width=0.43\textwidth]{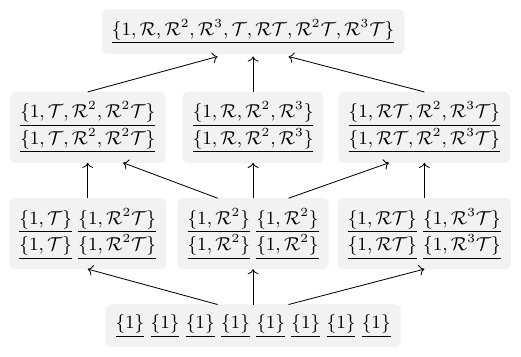}}
    \caption{Hasse diagram of the lattice of conjugate subgroups for the (a) abelian $D_4 \cong \ztzt$ and (b) non-abelian $D_8 \cong \bZ_4 \rtimes \ztwo$ groups.  The arrows represent an inclusion map, and two entries connected by a sequence of maps represent Landau-compatible phases. Horizontal lines represent the various vacua over which their residual symmetries are listed. We see that the phase with all symmetries broken (bottom) and all symmetries preserved (top) are Landau-compatible with all other phases.}
    \label{fig:Hasse}
\end{figure}
We present a discussion of spontaneous symmetry breaking (SSB) and Landau compatibility. For simplicity, we will assume that we are working with a classical statistical mechanical system whose symmetries $g \in G$ form a finite group of order $|G|$. 

\medskip

\noindent \emph{\textbf{Classifying SSB phases:}} A system is said to be in an SSB phase if it has multiple vacua that are not invariant under the full set of symmetries. For a given vacuum $v_\alpha$, let us denote {by} $H_\alpha \subset G$  the subgroup of residual symmetries that leaves it invariant. The set of symmetries in $G$, but not in $H_\alpha$, denoted $G\setminus H_{\alpha}$, transform the vacuum  $v_\alpha$ to other vacua $v_\beta$; although several elements of $G\setminus H_{\alpha}$ can give the same $v_\beta$. It is straightforward to see that
~~
\onecolumngrid
\begin{figure*}[t!]
    \includegraphics[width=\textwidth]{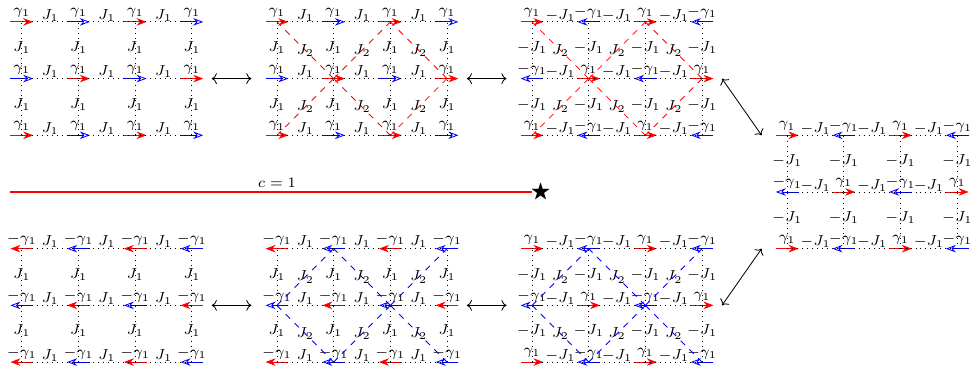}
    \caption{A path for bipartite lattices avoiding the unnecessary critical surface (denoted by $c=1$). We expect this surface to abruptly terminate.  \label{fig:UC_bipartite}}
\end{figure*}
~
\twocolumngrid
\noindent   the set of unique transformations are labelled by the cosets of $H_{\alpha}$ and starting with $v_\alpha$, the number of distinct vacua we can reach this way is given by the index $[G:H_\alpha]$. If $g_{\alpha\beta}$ takes $ v_\alpha\rightarrow v_\beta$, the residual symmetry group of $v_\beta$ is $H^{}_\beta = g_{\alpha\beta}^{} H_\alpha^{} g_{\alpha\beta}^{-1}$. The cosets of $H_\alpha$ and $H_\beta$ are identical. Thus, the generated $[G:H_\alpha]$ vacua family would be the same, independent of the initial choice of $v_\alpha$. We conclude that given a system with symmetry $G$, distinct SSB phases are labelled by distinct families of conjugate subgroups~\footnote{The identification by conjugation is often not stated in literature~\cite{ChenGuWen_1dSPTPhysRevB.84.235128}. However, it is important for correct classification, and without it we would overcount the number of SSB phases. For abelian groups, conjugation acts trivially, and various subgroups indeed represent distinct phases. For non-abelian groups, this is not the case.}.

For a finite group $G$, the different SSB phases are nicely organized by the lattice of conjugate subgroups and visualized by a Hasse diagram where the families of conjugate subgroups are connected by the presence of an inclusion map, i.e., when one family is a subgroup of the other. In \cref{fig:Hasse}, we have shown this for the dihedral symmetries $D_{2Q} \cong \bZ_Q \rtimes \ztwo $ considered in the main text for $Q=2,4$, with the presentation
\begin{align}
     D_{2Q} = \langle \cR, \cT| \cR^Q = \cT^2 = 1,~\cR\cT = \cT \cR^{-1}  \rangle \label{appeq:D2Q}.
\end{align}

\noindent \emph{\textbf{Landau compatibility:}} We can distinguish between Landau-compatible and incompatible transitions using the lattice of conjugate subgroups and its Hasse diagram. Two SSB phases represented by two families of conjugate subgroups are Landau-compatible if they are connected in the Hasse diagram by the composition of a sequence of arrows. Physically, we can understand this as follows: If we place ourselves in one of the vacua of the SSB phase with residual symmetries $H_\alpha \subset G$, we can treat it as a standalone system with symmetries $H_\alpha$. These can be spontaneously broken into a conjugate family of subgroups $K_{\alpha \beta} \subset H_\alpha$. All other transitions are Landau incompatible. Physically, a transition between Landau-incompatible phases cannot be understood by a hierarchical splitting of each vacuum, but rather by a more drastic process involving multiple vacua coming together and reorganizing themselves. 

In particular the fully symmetric phase with unique vacuum and fully broken phase with $|G|$ vacua are Landau compatible with all SSB phases. The interesting cases are the phases with partial symmetry breaking as we saw in the main text. 
Landau-compatible transitions are characterized by a change in the number of vacua (although this change does not guarantee compatibility). Moreover, Landau-incompatible transitions can occur between two SSB phases with the same number of vacua, as seen in the main text. 

\subsection*{\uline{Explicit paths avoiding the unnecessary critical surface for odd $Q$}}
In the main text, it was argued using symmetry that, for odd $Q$, the p-SSB$_\pm$ regions belong to the same phase and can be connected without encountering any phase transitions or violating any symmetries. Here we construct explicit such paths inspired by the so-called domain-wall pump~\cite{HermeleCMSATalk_2021}. 

\medskip
\noindent \emph{\textbf{Bipartite lattice:}} First, we consider a path, schematically shown in \cref{fig:UC_bipartite}, that works on bipartite lattices within an extended family of Hamiltonians of the form
\begin{multline}
	H = -\sum_{\moy{j,k}} J_1 \cos \left(\theta_j - \theta_k \right) -\sum_{\moy{\moy{j,k}}} J_2[j] \cos \left(\theta_j - \theta_k \right)  \\- \sum_{\ell = 1,2,\ldots}  \sum_j \gamma_\ell[j]  \cos \left(\ell Q \theta_j \right) \label{eq:Hclassical_bipartite}.
\end{multline}
\cref{eq:Hclassical_bipartite} contains nearest neighbour $\moy{j,k}$ and next nearest neighbour $\moy{\moy{j,k}}$ XY couplings that are allowed to differ on the two sublattices, which we label as red and blue. For our path, we begin with $\gamma_1>0$ in the p-SSB$_+$ region in one of its vacua, e.g.: $\moy{\theta_j} =0$ as shown in \cref{fig:UC_bipartite}. We want to flip the sign of $\gamma_1$ without encountering a phase transition. To achieve this, we first introduce a large $J_2>>J_1$ coupling only on the red sublattice which preserves the vacuum. We then flip the on-site anisotropy $\gamma_1 \rightarrow -\gamma_1$ on the blue sublattice as well as all $J_1 \rightarrow - J_1$; following which we remove the red $J_2 \rightarrow 0$. This produces an antiferromagnetic alignment between neighbouring spins and favours $\moy{\theta_j} = \pi$ on the blue spins. We then repeat the same steps for the other sublattice as shown in \cref{fig:UC_bipartite}. In the end, we obtain \cref{eq:Hclassical_bipartite} with $\gamma_1$ reversed and $\moy{\theta_j} = \pi$ on all spins, landing us on p-SSB$_-$. 

At each step of this symmetry-preserving path, we did not change the number of vacua. The magnitude of the order parameter detecting the p-SSB$_\pm$ phases, $\mathcal{O}$ defined in \cref{eq:SSBodd order parameters} has a nonvanishing average throughout the path although its sign develops a spatial variation. 

\begin{figure}[]
    \centering
    \includegraphics[width=.6\linewidth]{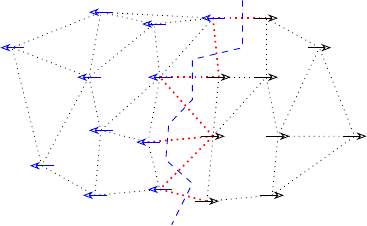}
    \caption{Sweeping a domain across the lattice transforms between p-SSB$_\pm$ without any phase transitions.}
    \label{fig:UC_Domain}
\end{figure}

\smallskip

\noindent \emph{\textbf{Any lattice:}} An unnecessary criticality avoiding path can be obtained for any lattice by nucleating a domain and sweeping it across the entire lattice as shown in \cref{fig:UC_Domain}. A domain is enclosed by a path on the dual lattice, by flipping all anisotropies, $\gamma_1 \mapsto -\gamma_1$, within and $J_1 \mapsto -J_1$, on the domain wall as shown in \cref{fig:UC_Domain}. Alternatively, instead of a single location, domains can be nucleated on various well-separated locations, grown and merged. 

For odd $Q$ the above paths does not risk a phase transition as the process of producing $\gamma_1$ of different signs between neighbouring spins $\moy{j,k}$ does not frustrate flipping the $J_1$ connecting them--- both favour a spin mismatch of $\theta_j - \theta_k = \pi$. This is not true for even $Q$.
\end{appendices}

\onecolumngrid

  \renewcommand\appendixpagename{\centering \Large\uline{ Supplemental materials}}
  \newpage
  
\begin{appendices}
\renewcommand{\thesection}{\Roman{section}}

\section{More details of the phase diagrams}\label{app:phasediagrams}
\begin{figure}[!ht]
     \centering
     \includegraphics[width=0.3\textwidth]{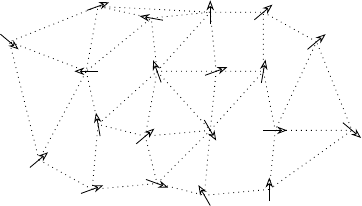}
     \caption{Schematic representation of a configuration for the JKKN model in \cref{appeq:Hclassical}. }
     \label{fig:lattice}
 \end{figure}

In this section, we provide more details on the phase diagrams for the classical Hamiltonians considered in the main text:
	\begin{align}
		H = -\sum_{\langle j,k \rangle} \cos \left(\theta_j - \theta_k \right) - \sum_{\ell = 1,2,\ldots} \gamma_\ell \sum_j  \cos \left(\ell Q \theta_j \right) \label{appeq:Hclassical}.
	\end{align}
The important symmetries form the group $D_{2Q} \cong \bZ_Q \rtimes \ztwo$ and have the following action on the planar rotors: 
\begin{align}
 \cR :\theta_j \mapsto \theta_j + \frac{2 \pi}{Q} ,\qquad ~\cT: \theta_j \mapsto -\theta_j. \label{appeq:symmetries}
\end{align}
The phase diagrams for each $Q$ are produced by replacing \cref{appeq:Hclassical} with the effective continuum gaussian theory in either of the two dual forms
 	\begin{align}
		S \approx  \int d^2x \left[ \frac{ \left( \nabla \phi\right)^2}{8 \pi^2 \beta}   - h  \cos (\phi)  -  \sum_{\ell}  \gamma_\ell \cos \left(\ell Q \theta\right)  \right] \leftrightarrow \int d^2x \left[ \frac{\beta }{2} \left( \nabla \theta\right)^2   - h  \cos (\phi)  -  \sum_{\ell}  \gamma_\ell \cos \left(\ell Q \theta\right)  \right], \label{appeq:S}
	\end{align}
and tracking the scaling dimensions of operators $\cos (\phi)$ and $\cos(Q \ell \theta)$. Let us begin by setting all $\gamma_\ell=0$, giving the familiar isotropic XY model in \cref{appeq:Hclassical}, whose phase diagram is reproduced by the gaussian model perturbed by a single operator $\cos(\phi)$ in \cref{appeq:S}.  It is well known that there exists a critical temperature $T_\gamma$ such that for $T>T_\gamma$, $\cos(\phi)$ is relevant and drives the system to the trivial, disordered phase and for $T<T_\gamma$, it is irrelevant and \cref{appeq:S} flows to the critical phase with pseudo long-range order described by the following effective theory
\begin{align}
		S \rightarrow \int d^2x  \frac{ \left( \nabla \phi\right)^2}{8 \pi^2 K_{\rm eff}}   \leftrightarrow \int d^2x \frac{ K_{\rm eff}}{2} \left( \nabla \theta\right)^2   . \label{appeq:Seff}
\end{align}
In this case, we can parametrize the theory using a single parameter $K_{\rm eff}$ which is related to the scaling dimension of $\cos (\phi)$. This determines all other scaling dimensions as well as correlation functions~\cite{JoseKirkpatrickPhysRevB.16.1217}
\begin{align}
   &[\cos (\phi)] = \pi K_{\rm eff},~ [\cos (\ell Q \theta)] = \frac{\ell^2 Q^2}{4 \pi K_{\rm eff}} , \label{eq:Keff_scaling}\\
   &\langle \cos(\phi(x)) \cos(\phi(y)) \rangle \sim |x-y|^{-2 \pi K_{\rm eff}},~\langle \cos(\ell Q \theta(x)) \cos(\ell Q \theta(y)) \rangle \sim |x-y|^{-\frac{\ell^2 Q^2}{2 \pi K_{\rm eff}}}.
\end{align} 
Let us now introduce $\gamma_\ell \neq 0$. For small values, their renormalization group (RG) flow equations are also determined from $K_{\rm eff}$ as follows
\begin{equation}
    \frac{d \gamma_\ell}{ds} \approx \left(2- \frac{\ell^2 Q^2}{4\pi K_{\rm eff}}  \right) \gamma_\ell,
\end{equation}
where $s$ is the length-scale of the RG flow. We see that when $\frac{\ell^2 Q^2}{4\pi K_{\rm eff}} <2$, $\gamma_\ell$ is relevant and grows at large distances whereas for $\frac{\ell^2 Q^2}{4\pi K_{\rm eff}} > 2$, $\gamma_\ell$ is irrelevant and shrinks. When multiple $\gamma_\ell$ are relevant, so long as they are all of the same order, the operators with the smallest scaling dimensions dominate (in this case corresponding to smallest  $\ell$). Since we tune $\gamma_1$ keeping all other $\gamma_\ell$ fixed, we only need to keep track of $\ell = 1,2$ to determine our phase diagram of interest. The various phases present in the model and their regimes are:
\medskip 

\begin{enumerate}
\itemsep0.5em
    \item A trivial disordered paramagnet driven by $\cos (\phi)$.
    \item Partial symmetry-breaking regions (p-SSB$_\pm$) when $\cos( Q \theta)$ is relevant, driven by $\gamma_1 \cos( Q \theta)$ for $\gamma_1>0$ and $\gamma_1 <0$ with vacua
	\begin{align}
		\vartheta^+_n = \frac{2 \pi n}{Q},~    \vartheta^-_n = \frac{(2n+1) \pi }{Q}.  \label{appeq:Qvacua}
	\end{align}
 \item Full symmetry-breaking phase (f-SSB)  when $\cos( 2Q \theta)$ is relevant, driven by $\gamma_2 \cos (2 Q \theta)$ for $\gamma_2 <0$ with vacua
 	\begin{align}
		\chi_n = \frac{\left(2n +1 \right) \pi}{2Q},~\text{for } n=1,2,\ldots,2Q \label{appeq:2Qvacua}.
	\end{align}
 \item A gapless phase when all operators $\cos (\phi)$ and $\cos( \ell Q \theta)$ are irrelevant.
\end{enumerate}
\medskip 
Along the $\gamma_1=0$ line, it is also useful to define the following critical temperatures: 
\medskip 
\begin{itemize}
    \itemsep0.5em 
    \item $T^\alpha_Q$, when $\cos(2 Q \theta)$ is marginal, i.e., $K_{\rm eff} = Q^2/(2\pi) $,
    \item $T^\beta_Q$ when $\cos (Q \theta)$ is marginal, i.e., $K_{\rm eff} = Q^2/(8\pi)$ and 
    \item  $T^\gamma$ when $\cos (\phi)$ is marginal, i.e., $K_{\rm eff} = 2/\pi$. 
\end{itemize}
\medskip 
For $Q\ge3$, along the $\gamma_1=0$ line, for the range of temperatures between $T_\alpha^Q$ and $T^\beta_Q$, we get the Landau-incompatible transitions (even $Q$) and unnecessary critical line (odd $Q$) when $\cos(Q \theta)$ is the only relevant operator. 

We may ask if there are other perturbations missed in the models we have considered that can qualitatively change the phase diagram. It is easy to verify that there are no other \emph{primary}~\cite{Ginsparg1988applied} scaling operators that are allowed by symmetry. For example, $\sin(Q\theta)$, which could connect the $\gamma_1<0$ and $\gamma_1 >0$ p-SSB$_\pm$ regions, is  disallowed by the $\theta \mapsto -\theta$ symmetry. But what about \emph{descendant} operators~\cite{Ginsparg1988applied}? In particular for $Q\ge 3$, for $T^\alpha < T < \min(T^\beta_Q,T^\gamma)$ the only relevant symmetry-allowed primary operator in \cref{appeq:S} is $\cos(Q \theta)$. Discounting the symmetry-disallowed operators, we should consider $\partial_x\left( \cos(Q\theta)\right)$ with dimension $[\cos (Q\theta)]+1$. This is symmetry-allowed and would be relevant when $[\cos (Q\theta)] < 1$. 
However, this operator is a boundary term and would not affect the bulk phase diagram. In this work, we do not explore the interesting question of how the presence of descendants can affect boundary phenomena. 

The phase diagram of \cref{appeq:Hclassical} is also qualitatively reproduced by the quantum analogues with Hamiltonians of the form
	\begin{align}
		H &= -H_{\textrm{XXZ}} - h H_0 - \sum_{\ell = 1,2,\ldots} \gamma_{\ell} H_{ \ell Q}  \label{appeq:Hquantum} \\
		\Bigg(\text{Where}\quad H_{\text{XXZ}} &=  \sum_j \left(  S^x_j S^x_{j+1} + S^y_j S^y_{j+1}+ \Delta  S^z_j S^z_{j+1}  \right) ,\qquad
		H_{\ell Q} = \sum_j \left( \prod_{l=0}^{\ell Q} S^+_{j+l} + \prod_{l=0}^{\ell Q} S^-_{j+l} \right), \nonumber \\\qquad\text{ and }\quad
		H_0 &=\sum_j (-1)^j S^z_j \qquad\text{ or }\qquad \sum_j (-1)^j \left(  S^x_j S^x_{j+1} + S^y_j S^y_{j+1} \right)\Bigg). \nonumber 
	\end{align}
 for $|\Delta| < 1$ by identifying $K_{\rm eff} = 2 \arccos \Delta$ via Bethe ansatz~\cite{Haldane_BetheLL_1981153}.

\subsection{Double frequency sine-Gordon and f-SSB to \texorpdfstring{p-SSB$_\pm$}{p-SSB±} Ising transition}
\label{sec:MultifSG}
\begin{figure}[!ht]
    \centering
    \subfloat[$\gamma_1 \rightarrow \infty$]{\includegraphics[width=0.3\textwidth]{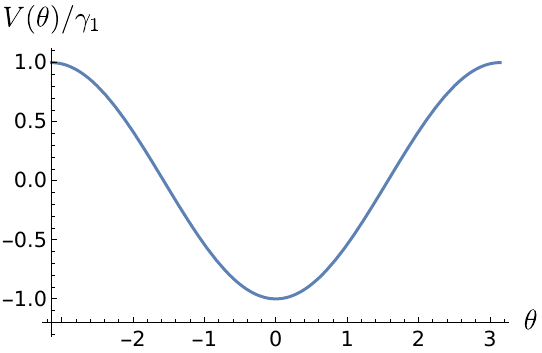}}
       \subfloat[$\gamma_1 = 4 |\gamma_2|$]{\includegraphics[width=0.3\textwidth]{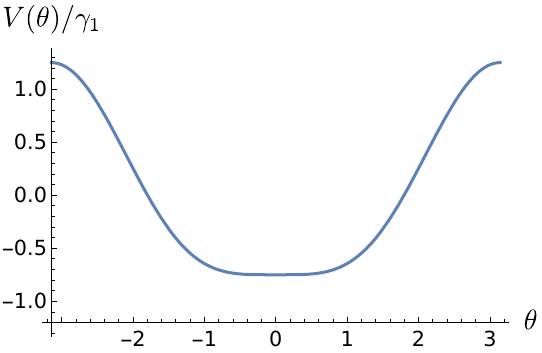}}
       \subfloat[$\gamma_1 = |\gamma_2|$]{\includegraphics[width=0.3\textwidth]{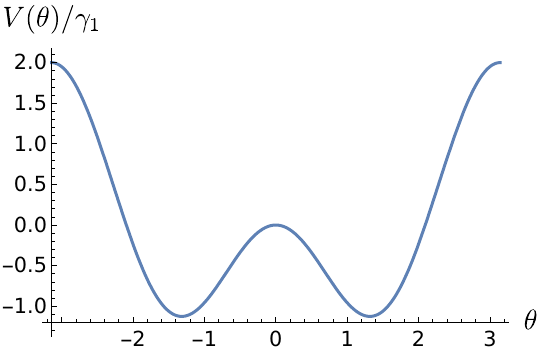}}
    \caption{The Ising transition for $\gamma_1>0$ seen from the potential ~\cref{appeq:MultifqSG_potential} in the double-frequency sine-Gordon model in \cref{appeq:S_doublefqSG}. We see a transition at $\gamma_1 = 4 |\gamma_2|$ .}
    \label{fig:MultifqSG}
\end{figure}
The transition between the  p-SSB$_\pm$ and f-SSB phases, dominated by $\cos(Q\theta)$ and $\cos(2Q \theta)$ respectively, belongs to the Ising universality class. To understand why, let us focus on $Q=1$ where for $T < T^\gamma$, $\cos (\phi)$ is irrelevant, while $\cos(\theta)$ and $\cos (2\theta)$ are relevant. Thus, the action~\cref{appeq:S} reduces to
\begin{align}
    S \approx  \int d^2x \left[ \frac{\beta }{2} \left( \nabla \theta\right)^2   -  \gamma_1  \cos(\theta) -\gamma_2 \cos (2\theta)  \right]. \label{appeq:S_doublefqSG}
\end{align}
This is the so-called double-frequency Sine-Gordon model~\cite{DELFINO1998675} whose analysis reveals the Ising transition. Let us understand this by looking at the potential
\begin{align}
    V(\theta) = -\left( \gamma_1  \cos(\theta) +\gamma_2 \cos( 2\theta)  \right) \label{appeq:MultifqSG_potential}
\end{align}
For $\gamma_2 <0$, $V(\theta)$ has a unique minimum for large $|\gamma_1|$ at $\theta = 0$ (for $\gamma_1>0$) and $\theta = \pi$ (for $\gamma_1 <0$). As we reduce $|\gamma_1|$, we see in \cref{fig:MultifqSG} that the $\theta \mapsto -\theta$ symmetry is spontaneously broken in the vacuum at $\gamma_1 = 4 |\gamma_2|$. Near this point, we can Taylor expand \cref{appeq:S_doublefqSG} to get (after appropriate redefinitions) a 2d real scalar field theory with Ising symmetry
\begin{align}
    S \approx \int d^2x \left[ \frac{(\nabla \theta)^2}{2}  + \frac{m^2}{2} \theta^2 + \lambda \theta^4 + \ldots \right]\label{appeq:S_realscalar} ~.
\end{align}
This flows to the Ising universality class at criticality i.e $m \rightarrow 0$. The same story can be repeated for any $Q$ by replacing $\theta \rightarrow Q \theta$ in \cref{appeq:S_doublefqSG,appeq:MultifqSG_potential}. For large $|\gamma_1|$, we will now get $Q$ minima, but by Taylor expanding around each minimum, we get \cref{appeq:S_realscalar}.

\subsection{Self-dual sine-Gordon model and PM to \texorpdfstring{p-SSB$_\pm$}{p-SSB±} transition}
\label{sec:SDSG}
To understand the transition from the symmetry preserving PM and the p-SSB$_\pm$ phases dominated by $\cos (\phi)$ and $\cos (Q \theta)$ respectively, we want to focus on the situation when the two operators have the same scaling dimension. This happens when $[\cos (\phi)]  = [\cos (Q \theta)] = \frac{Q}{2}$. We see that the both operators are not irrelevant for $Q \le 4$ when a direct transition can exist. As explained in Ref.\cite{LECHEMINANT_SelfDualSineGordon_2002502}, this is described by the self-dual Sine-Gordon model
\begin{align}
    S = \int d^2x \left[ \frac{ \left( \nabla \phi\right)^2}{4 \pi Q}   - g\left(  \cos (\phi) + \cos (Q \theta) \right) \right]. \label{appeq:SDSG_Q}
\end{align}
 For $Q \le 4$, when $\cos (\phi)$ and $\cos (Q \theta)$ are both not irrelevant, the fate of \cref{appeq:SDSG_Q} under RG flow was analyzed in Ref.~\cite{LECHEMINANT_SelfDualSineGordon_2002502}. Their results are summarized as follows:
 \medskip 
\begin{enumerate}
\itemsep1em
    \item For $Q=1$, \cref{appeq:SDSG_Q} flows to a trivial gapped state. Thus, there is no phase transition between the $\cos (\phi)$ and $\cos(\theta)$ dominant regions and they are smoothly connected.
    \item For $Q=2$, \cref{appeq:SDSG_Q} flows to the Ising universality class.
    \item For $Q=3$, \cref{appeq:SDSG_Q} flows to the 3-state Potts universality class.
    \item For $Q=4$, $\cos (\phi)$ and $\cos( Q \theta)$ are both marginal and induce a flow in the $c=1$ conformal manifold~\cite{DijkgraafVerlindeVerlinde}, along the orbifold branch which describes the scaling limit of the Ashkin-Teller model~\cite{AshkinTellerPhysRev.64.178}. Thus, the transition has varying critical exponents.
\end{enumerate}
\medskip 
 The  RG flow of \cref{appeq:SDSG_Q} is trivial for $Q\ge 5$ when $\cos (\phi)$ and $\cos( Q \theta)$ are irrelevant, giving us a $c=1$ gaussian theory. This represents an intermediate gapless phase rather than a direct transition between PM and p-SSB$_\pm$ phases.

\subsection{Phase diagrams for \texorpdfstring{$\gamma_2>0$}{γ2>0}}
In the main text, we presented phase diagrams for  $\gamma_2 <0$. Here, we discuss the same for $\gamma_2>0$. The main difference is in the low-temperature regime. Whereas for $\gamma_2 <0$ we saw the f-SSB phase with all symmetries broken, for $\gamma_2 >0$ we see that the partial symmetry-breaking regions p-SSB$_{\pm}$ persist but are now separated by a first-order line for $T < T^\alpha_Q$ where the vacua of both the p-SSB$_{\pm}$ regions coexist as seen by minimizing $\left(-\cos(2Q \theta)\right)$. We now combine all this to obtain the phase diagrams for various $Q$.

\subsection{\texorpdfstring{$Q=1$}{Q=1} phase diagrams}
\begin{figure}[!ht]
    \centering
    \subfloat[$\gamma_2<0$]{\includegraphics[]{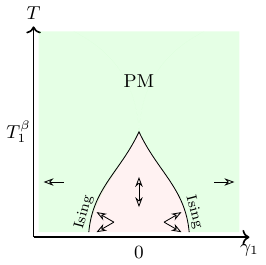}}
    \subfloat[$\gamma_2=0,~\gamma_3>0$]{\includegraphics[]{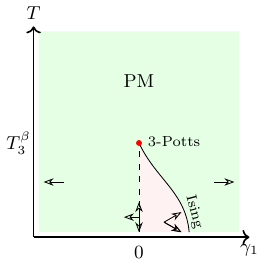}}
    \subfloat[$\gamma_2>0$]{\includegraphics[]{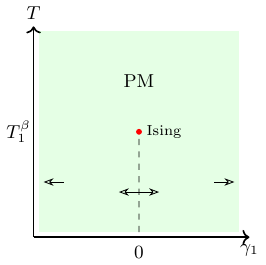}
  }
    \caption{Phase diagrams for the $Q=1$ Hamiltonian of~\cref{appeq:Hclassical}. Arrowheads indicate the pinned values of $\moy{\theta}$ . First and second order transitions are indicated by solid and broken lines.}
    \label{fig:Q1}
\end{figure}
Let us begin with the trivial, yet instructive case of the phase diagram for $Q=1$. The only non-trivial symmetry is $\cT$ in \cref{appeq:symmetries} and generates a $\ztwo$ group
\begin{align}
    \ztwo \cong \{1, \cT \}.
\end{align}
An interesting observation is the absence of a direct phase transition between the $\cos(\theta)$ and $\cos (\phi)$ dominant regions. This was argued from the analysis of the RG flow of the self-dual sine-Gordon model in \cref{appeq:SDSG_Q} with $Q=1$. A much simpler way to see the same fact is to consider the $Q=1$ quantum Hamiltonian in~\cref{appeq:Hquantum}
\begin{align}
    H = -H_{\textrm XXZ} - h \sum_j (-1)^j S^z_j - \gamma_1 \sum_j S^x_j + \ldots ~.
\end{align}
By setting $\gamma_1 \rightarrow \infty$, we get the $\cos(\theta)$ dominant phase with a product ground state. This is clearly adiabatically connected to the ground state of $h \sum_j (-1)^j S^z_j$ which is equivalent to the $\cos (\phi)$ dominated high temperature phase of the classical model.

We see that the nature of the phase diagram depends on the sign of $\gamma_2$, as discussed before and shown in \cref{fig:Q1}(a,c). For $\gamma_2 = 0$, the phase diagram is modified as shown in \cref{fig:Q1}(b). The Ising and first order lines can be analyzed using the double-frequency sine-Gordon potential
\begin{align}
    V(\theta)= - \left(\gamma_1 \cos(\theta) + \gamma_3 \cos(3 \theta) \right)
\end{align}
and repeating the analysis in \cref{sec:MultifSG}. 
The 3-state Potts universality class where the two lines meet at $T^\beta_3$ and $\gamma_1 = 0$ is obtained from the flow of the self-dual sine-Gordon model in \cref{appeq:SDSG_Q} with $Q=3$.

If we further tune $\gamma_3 = 0$ (not shown in \cref{fig:Q1}),  we get a continuous unnecessary critical line in the phase diagram for $T< T^\gamma$ along $\gamma_1 = 0$. This is not stable and needs fine-tuning three parameters, unlike for $Q \ge 3$ where the unnecessary criticality needs fine-tuning only one parameter, i.e, $\gamma_1 = 0$.

\subsection{\texorpdfstring{$Q=2$}{Q=2} phase diagrams}
\begin{figure}[!ht]
    \centering
    \subfloat[$\gamma_2 <0$]{\includegraphics[]{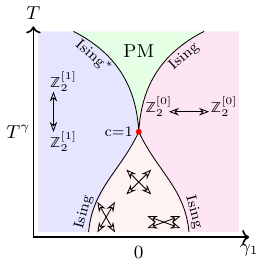}}
      \subfloat[$\gamma_2 =0$]{\includegraphics[]{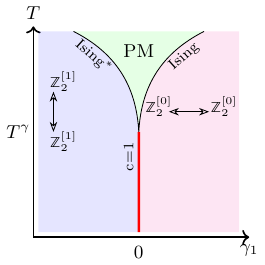}}
      \subfloat[$\gamma_2 >0$]{\includegraphics[]{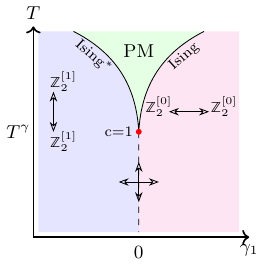} }
    \caption{Phase diagrams for the $Q=2$ Hamiltonian of ~\cref{appeq:Hclassical}.  Arrowheads indicate the pinned values of $\moy{\theta}$ . First and second order transitions are indicated by solid and broken lines. A direct second order transition between Landau-incompatible phases requires tuning two parameters.}
    \label{fig:Q2}
\end{figure}
We now consider the $Q=2$ model of \cref{appeq:Hclassical} whose phase diagrams are shown in \cref{fig:Q2}. The symmetry group of~\cref{appeq:symmetries} is abelian,
\begin{align}
 \ztwo \times \ztwo \cong \{1,\cR,\cT, \cR \cT \}. \label{appeq:Z2Z2}   
\end{align}
The phase diagrams for various $\gamma_2$ values are shown in \cref{fig:Q2}. We see that a direct Landau-incompatible transition is not stable, and needs fine-tuning two parameters for $\gamma_2\neq 0$, i.e., $\gamma_1 = 0$ and $T = T^\gamma$ or $\gamma_1 = \gamma_2 = 0$. However, this fine-tuning can occur accidentally quite naturally in quantum models with aesthetic appeal. For instance, setting $\gamma_{\ell \ge 2} = 0$ in \cref{appeq:Hquantum}  yields the XYZ model~\cite{LSMOG} with a staggered magnetic field, 
\begin{align}
    H = -(1-2\gamma_1) \sum_j S^x_j S^x_{j+1} -(1+2\gamma_1) \sum_j S^y_j S^y_{j+1} -\Delta \sum_j S^z_j S^z_{j+1} -h \sum_j (-1)^jS^z_j.\label{appeq:XYZ}
\end{align}
The phase diagram of \cref{appeq:XYZ} exhibits a direct Landau-incompatible transition. However, this is special to the specific model and unstable to symmetric four-body perturbations introduced by $\ell  =2$ in \cref{appeq:Hquantum}.

A final interesting feature of \cref{fig:Q2} is the presence of symmetry-enriched Ising criticality~\cite{VerresenSEC,AP_SEC_PhysRevB.108.245135}. Transitions between PM and p-SSB$_\pm$ phases belong to the Ising universality class. However, symmetries act on the two branches in different ways. This is seen from the fact that the order parameters that correspond to the primaries of the Ising CFT transform as different irreducible representations of the $\ztwo \times \ztwo$ symmetry in \cref{appeq:Z2Z2}. It is sufficient to track the primary field $\sigma$. For the transition to p-SSB$_+$, this corresponds to the lattice operator $\sigma \sim \cos(\theta)$ which is charged under $\cR$ but not $\cT$, while for the transition to p-SSB$_-$, this corresponds to $\sigma \sim \sin(\theta)$ which carries both $\cR$ and $\cT$ charges. The distinct charge assignments mean that the two Ising CFT branches cannot be connected trivially. In \cref{fig:Q2}, they pass through a different universality class with $c=1$.

\subsection{\texorpdfstring{$Q=3$}{Q=3} phase diagrams}
\begin{figure}[!ht]
    \centering
    \subfloat[$\gamma_2 = 0$]{\includegraphics[]{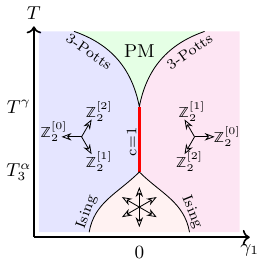}}
  \subfloat[$\gamma_2 >0$]{\includegraphics[]{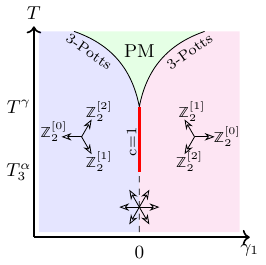}}
    \caption{Phase diagrams for the $Q=3$ Hamiltonian of ~\cref{appeq:Hclassical}.  Arrowheads indicate the pinned values of $\moy{\theta}$ . First and second order transitions are indicated by solid and broken lines.}
    \label{fig:Q3}
\end{figure}
The phase diagrams for the $Q=3$ Hamiltonian are shown in \cref{fig:Q3}. The two p-SSB$_\pm$ regions correspond to the same phase and the two branches of transitions between them and the PM belonging to the 3-state Potts universality class. 

\subsection{\texorpdfstring{$Q=4$}{Q=4} phase diagrams}
\begin{figure}
    \centering
    \subfloat[$\gamma_1 <0$]{\includegraphics[]{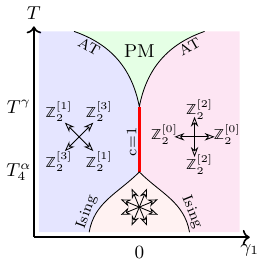}}
     \subfloat[$\gamma_1 >0$]{\includegraphics[]{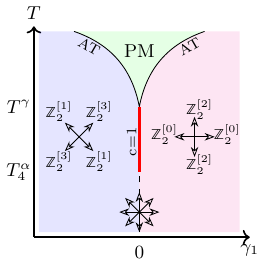}}
  \caption{Phase diagrams for the $Q=4$ Hamiltonian of~\cref{appeq:Hclassical}.  Arrowheads indicate the pinned values of $\moy{\theta}$. First- and second-order transitions are indicated by solid and broken lines. The lines marked `AT' have varying critical exponents and correspond to the orbifold branch of the conformal manifold that describes the critical Ashkin-Teller model.}
    \label{fig:Q4}
\end{figure}
The phase diagram for $Q=4$ also presents an interesting case. This is the largest $Q$ in which there is a direct transition between the PM and p-SSB$_\pm$ phases. This is described by the self-dual sine-Gordon model in \cref{appeq:SDSG_Q} for $Q=4$. The two operators in \cref{appeq:SDSG_Q}, i.e., $\cos (\phi)$ and $\cos(Q\theta)$ with the same scaling dimensions are relevant for $Q=1,2,3$. For $Q=4$, however, they are \emph{marginal} and tuning their coefficient induces a flow on the conformal manifold with central charge $c=1$ along the so-called orbifold branch~\cite{DijkgraafVerlindeVerlinde}. This branch also describes the critical Ashkin-Teller model~\cite{AshkinTellerPhysRev.64.178} and has varying critical exponents, similar to the $c=1$ compact boson branch that describes the direct transition between the p-SSB$_\pm$ phases. It is known that the orbifold and compact boson branches meet at the Kosterlitz-Thouless transition which occurs at $T =T^\gamma$~\cite{DijkgraafVerlindeVerlinde} for our model.

The two lines of transitions between PM and p-SSB$_\pm$ seem to have different symmetry enrichments seen by tracking the symmetry charges of the order parameters for SSB$_\pm$, $\cos(2\theta)$ and $\sin(2\theta)$. But, as explained in Ref.~\cite{VerresenSEC} and seen in \cref{fig:Q4}, they are smoothly connected via the KT point without leaving the conformal manifold and should nominally be considered as belonging to the same symmetry-enriched class.

\subsection{\texorpdfstring{$Q\ge 5$}{Q≥5} phase diagrams}
\begin{figure}[!ht]
    \centering
    \subfloat[$\gamma_1<0$]{		
    \includegraphics[]{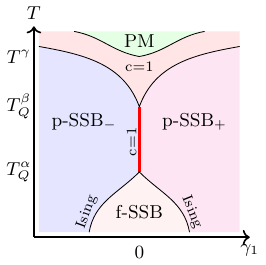}}
    \subfloat[$\gamma_1>0$]{		
        \includegraphics[]{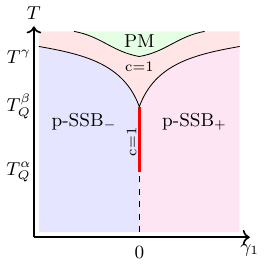}}
         \caption{Phase diagrams for the $Q\ge 5$ Hamiltonians of~\cref{appeq:Hclassical}.   First- and second-order transitions are indicated by solid and broken lines.}
\end{figure}

Finally, we consider the remaining cases $Q \ge 5$, where a direct transition between the PM and p-SSB$_\pm$ does not exist. Rather there is an intermediate gapless phase. In this region, all symmetry-allowed operators in \cref{appeq:S}, i.e., $\cos(\ell Q \theta)$ and $\cos (\phi)$, are irrelevant. The gapless phase terminates at low temperatures when $\cos (Q \theta)$ becomes marginal and at high temperatures when $\cos (\phi)$ becomes marginal. Along the $\gamma_1 = 0$ line, these occur at $T = T^\beta_Q$ and $T = T^\gamma$ respectively. As we increase $Q$, the size of the p-SSB$_\pm$ regions shrinks, whereas the gapless phase grows. In the formal limit of $Q \rightarrow \infty$, $\cR$ becomes a full $U(1)$ symmetry, and we recover the phase diagram of the ordinary classical XY model.

\section{Comments on anomalies}

The study of anomalous symmetries has a long history in high energy and condensed matter physics in different contexts, which got a fillip with the study of topological phases. Anomalies are non-trivial manifestations of symmetries and are characterized by different properties:
\medskip
\begin{enumerate}
\itemsep0.5em
    \item They disallow a strictly on-site representation~\cite{ElseNayakCohomologyPhysRevB.90.235137}.
    \item They present an obstruction to gauging~\cite{WenAnomalies_PhysRevD.88.045013,fradkin2021quantum}.
    \item They disallow a trivial symmetry-preserving phase~\cite{MetlitskiThorngrenDQC_PhysRevB.98.085140,AP_LSM2020elementary} in compatible phase diagrams.
\end{enumerate}
\medskip
These properties are believed to be equivalent, although to the best of our knowledge, this has not been proven yet. 

\subsection{Anomalies in quantum systems}
In quantum systems, anomalies can manifest themselves in two distinct ways. The first case is where the microscopic symmetries are anomalous on the full Hilbert space, as in the case of (a) Lieb-Schultz-Mattis (LSM) constrained spin systems~\cite{LSMOG,Hastings_LSM_PhysRevB.69.104431,Oshikawa_LSM_PhysRevLett.84.1535,AP_LSM2020elementary} where there exists a mixed anomaly between on-site projective representations and lattice symmetries, and (b) boundaries of symmetry protected topological (SPT) phases~\cite{LSMSPT_2018PhysRevB.97.054412_Xu,DQC_1D_SPT_Zhang_PhysRevLett.130.026801,LSM_AnomalySPT_2017PhysRevB.96.195105_Ryuetal} where symmetries are represented on boundary degrees of freedom as anomalous non-on-site finite depth circuits. These systems do not host a trivial symmetry-preserving phase anywhere in their phase diagrams. The second case is where the microscopic symmetries are not anomalous on the full second-quantized Hilbert space but are anomalous when restricted to certain symmetry sectors. This is the case for LSM constrained fermion systems where a trivial insulator is forbidden at certain fermion densities but not all. By changing the symmetry sector of the ground state, e.g., by tuning a chemical potential or magnetic field, a trivial phase can be obtained in the phase diagram. 

In both settings, the anomalies are kinematic and have a clear microscopic origin that can be diagnosed using microscopic probes~\cite{Oshikawa_LSM_PhysRevLett.84.1535,LSM_2020_ElseThrngrenPhysRevB.101.224437,LatticeHomotopyPhysRevLett.119.127202}. Often, it is convenient to employ an effective field-theoretic lens ~\cite{Wangetal_DQC_2017_PhysRevX.7.031051,MetlitskiThorngrenDQC_PhysRevB.98.085140,LSMAnomaly2023_ChengSeiberg} to understand anomalies using gauge fields. This is especially useful in tracking lattice symmetries which often `emanate' to an internal symmetry on the effective low-energy fields~\cite{Wangetal_DQC_2017_PhysRevX.7.031051,MetlitskiThorngrenDQC_PhysRevB.98.085140,LSMAnomaly2023_ChengSeiberg}. In particular, as shown in Ref~\cite{MetlitskiThorngrenDQC_PhysRevB.98.085140}, anomalies present only under restriction of symmetry sectors manifest themselves in the infrared by acting unfaithfully as a quotient. For instance, spinful fermions on a square lattice with an $SU(2)$ symmetry, when restricted to half-filling acts as $SO(3) \cong SU(2)/\ztwo$ which has an LSM anomaly. These emanant symmetries should be distinguished from \emph{emergent} ones. The latter are  without any microscopic origin and are broken by irrelevant operators, whereas the former are not. 

Deconfined criticality is sometimes associated to the microscopic symmetries being realised anomalously~\cite{Senthil2023DQCReview} either on the ultraviolet~\cite{LSMOG,Hastings_LSM_PhysRevB.69.104431,Oshikawa_LSM_PhysRevLett.84.1535,AP_LSM2020elementary} or infrared~\cite{Wangetal_DQC_2017_PhysRevX.7.031051,MetlitskiThorngrenDQC_PhysRevB.98.085140,LSMAnomaly2023_ChengSeiberg} degrees of freedom. Note that this prejudice is not universally held, as evidenced by the search for a direct DQC transition between the N\'{e}el and valence-bond solid (VBS) phases on the honeycomb lattice~\cite{KedarDQCHoneycombPhysRevLett.111.087203,MetlitskiThorngrenDQC_PhysRevB.98.085140} which is not constrained by the LSM theorem. However, recent work~\cite{HeFuzzyDQC} has indicated that  this transition may in fact not exist. Our work proves that even if this particular DQC is not present in the honeycomb lattice, there is no fundamental obstruction to observing Landau-incompatible transitions on the honeycomb or other LSM-trivial lattices.

\subsection{Anomalies in classical systems}
A well known result is that the ground state phases of a $d$-dimensional quantum system can be reproduced by an appropriate $d$+1 dimensional classical system~\cite{Sachdev_book}. When the quantum system is anomalous, and its phase diagram does not contain a trivial, symmetric phase, how does this manifest in the corresponding classical system? In particular, at high enough temperatures, classical systems are expected to restore all symmetries and produce a trivial disordered phase, seemingly in contradiction to the anomaly constraint of the corresponding quantum system~\footnote{We thank the anonymous referee for posing this interesting question}! The resolution of this paradox comes from the observation that the classical system corresponding to an anomalous quantum system can be constrained. For example, consider the XXZ spin chain which has a well-known LSM anomaly~\cite{MetlitskiThorngrenDQC_PhysRevB.98.085140}. Its corresponding classical system is the six-vertex model \cite{Baxter,Franchini17}. If we consider the partition function for this model with unit Boltzmann weights, representing the infinite temperature limit, we have
\begin{equation}
    \cZ_{6v} = \sum_{\mathcal{C}_{6v}} 1.
\end{equation}
This is in a disordered phase characterised by pseudo long-range order and algebraic correlations (it corresponds to the eight-vertex model at a critical temperature) \cite{Baxter}, and hence \emph{does not} correspond to the trivial disordered phase which has exponentially decaying correlations. The six-vertex constrained nature of the configuration space $\mathcal{C}_{6v}$ over which the partition function sum is performed introduces non-trivial correlations even when all Boltzmann weights are unity. This is also true of other constrained systems, for example, two-dimensional dimer models~\cite{chalker2017spin,kenyon2003introductiondimermodel}. On the other hand, if the configuration space were to be constraint-free, as is true of the classical Ising and XY model, the partition function with unit Boltzmann weights would indeed be trivial with exponentially decaying correlations. 
\end{appendices}

\end{document}